\documentclass{aa}  
\usepackage{graphicx}
\usepackage{txfonts}
\usepackage{natbib}
\usepackage{color}
\usepackage{soul}
\definecolor{stan}{rgb}{0,0,1}

\def\<<{{\ll}}
\def\>>{{\gg}}

\def\spose#1{\hbox to 0pt{#1\hss}}
\def\ltwig{\mathrel{\spose{\lower 3pt\hbox{$\mathchar"218$}}
     R_{\rm A}ise 2.0pt\hbox{$\mathchar"13C$}}}
\def\gtwig{\mathrel{\spose{\lower 3pt\hbox{$\mathchar"218$}}
     R_{\rm A}ise 2.0pt\hbox{$\mathchar"13E$}}}
\def\+/-{{\pm}}
\def\=={{\equiv}}

\def\Rstar{R_{\ast}}

\def\vth{v_{\rm th}}

\def\Game{\Gamma_{\rm e}}
\def\kape{\kappa_{\rm e}}
\def\kapo{\kappa_{\rm o}}
\def\kapmax{\kappa_{\rm max}}
\newcommand{\beq}{\begin{equation}}
\newcommand{\eeq}{\end{equation}}
\newcommand{\beqa}{\begin{eqnarray}}
\newcommand{\eeqa}{\end{eqnarray}}

\begin{document} 

   \title{2D wind clumping in hot, massive stars from hydrodynamical line-driven instability simulations using a pseudo-planar approach}

   \author{J.O. Sundqvist\inst{1}\and
     S.P. Owocki\inst{2}\and
     J. Puls\inst{3} }

   \institute{KU Leuven, Instituut voor Sterrenkunde, Celestijnenlaan 200D, 3001 Leuven, 
   Belgium\\ \email{jon.sundqvist@kuleuven.be}\and Department of Physics and Astronomy, Bartol Research Institute,
 University of Delaware, Newark, DE 19716, USA\and
     Universit\"atssternwarte M\"unchen, Scheinerstr. 1,
     81679 M\"unchen, Germany}
                                 
   \date{Received 2017-08-04; accepted 2017-10-20}

 
  \abstract
  {Clumping in the radiation-driven winds of hot, massive stars arises 
  naturally due to the strong, intrinsic instability of line-driving (the `LDI'). But LDI wind models have 
  so far mostly been limited to 1D, mainly because of severe computational challenges regarding 
  calculation of the multi-dimensional radiation force.}   
   {To simulate and examine the dynamics and multi-dimensional nature of wind structure resulting from the LDI.} 
   {We introduce a `pseudo-planar', `box-in-a-wind' method that allows us to efficiently compute the line-force in the radial
   and lateral directions, and then use this approach to carry out 2D radiation-hydrodynamical simulations of the time-dependent wind.}
   {Our 2D simulations show that the LDI first manifests itself by mimicking the typical shell-structure seen in 1D 
   models, but how these shells then quickly break up into complex 2D density and velocity structures, 
   characterized by small-scale density `clumps' embedded in larger regions of fast and rarefied gas. Key results of 
   the simulations are that density-variations in the well-developed wind statistically are quite isotropic and that characteristic 
   length-scales are small; a typical clump size is $\ell_{\rm cl}/R_\ast \sim 0.01$ at $2 R_\ast$, 
  thus resulting also in rather low typical clump-masses $m_{\rm cl} \sim 10^{17}$\,g. Overall, our results 
  agree well with the theoretical expectation that the characteristic scale for LDI-generated wind-structure 
  is of order the Sobolev length $\ell_{\rm Sob}$. We further confirm some earlier results that lateral `filling-in' of 
  radially compressed gas leads to somewhat lower clumping factors in 2D simulations than in 
  comparable 1D models. We conclude by discussing an extension of our method toward rotating LDI wind 
  models that exhibit an intriguing combination of large- and small-scale structure extending down to the wind base.}
{}
     \titlerunning{2D Wind Clumping}
   \maketitle
%

\section{Introduction}
\label{intro}

For massive, hot stars of spectral types OBA, scattering and absorption in spectral 
lines transfer momentum from the star's intense radiation field to the plasma, and so provide the force necessary to
overcome gravity and drive a strong stellar wind outflow \citep[see][for an extensive 
review]{Puls08}. The first quantitative description of such line-driving was given in the seminal 
paper by \citet{Castor75}: hereafter `CAK'. Like many wind models to date, 
CAK used the so-called Sobolev approximation \citep{Sobolev60} 
 to compute the radiative acceleration. This assumes that hydrodynamic flow 
 quantities\footnote{Or more specifically, occupation number densities and source functions.} are 
constant over a few Sobolev lengths $ \ell_{\rm Sob} = \varv_{\rm th}/(d\varv_{\rm n}/dn)$ 
(for ion thermal speed $\varv_{\rm th}$ and projected velocity gradient $d\varv_{\rm n}/dn$ along a coordinate 
direction $\hat{n}$), 
allowing then for a \textit{local} treatment of the line radiative transfer.  

Such a Sobolev approach ignores the strong `line deshadowing instability' (LDI)
that occurs on scales near and below the Sobolev length \citep{Owocki84}; 
numerical radiation-hydrodynamic modeling of the non-linear evolution of the 
LDI shows that the time-dependent wind develops a very inhomogeneous,    
`clumped' structure \citep{Owocki88, Feldmeier97, Dessart03, 
Dessart05, Sundqvist13, Sundqvist15}. Such clumpy LDI models 
provide a natural explanation for a number of observed 
phenomena in OB-stars, such as the soft X-ray emission and
broad X-ray lines observed by orbiting telescopes like {\sc chandra}
and {\sc xmm-newton} \citep{Feldmeier97, Berghofer97, Gudel09,
Cohen10, Martinez17}, the extended regions of zero residual flux typically
seen in saturated UV resonance lines \citep{Lucy83, Puls93,
Sundqvist10}, and the migrating spectral sub-peaks superimposed on
broad optical recombination lines \citep{Eversberg98, Dessart05b,
Lepine08}.

But a severe limitation of most of the above-mentioned models is their assumed spherical 
symmetry. The fact that most LDI simulations in the past have been limited to 1D is mainly a 
consequence of the computational cost associated with carrying out the non-local integrals 
needed to compute the radiation acceleration at each simulation time-step, 
while simultaneously resolving length-scales below $\ell_{\rm Sob}$. Specifically, following 
the general escape-integral methods 
developed by \citet{Owocki96}, some $n_{\rm x} \approx 3 \varv_\infty/\varv_{\rm th} 
\approx 1000$ discrete frequency points are typically needed to properly resolve 
line profiles and model the expanding flow. In 2D or 3D, a proper treatment of the 
multi-dimensional wind further requires integrations along a set of oblique rays in order 
to compute the radiative force in the radial and lateral directions. A major issue then 
becomes misalignment of nonradial rays with the discrete numerical 
grid (i.e. that oblique ray-integrations from any given point 
in the mesh in general do not intersect any other point), 
requiring that all integrations be repeated for each grid node (and 
also then involving complex interpolation schemes to trace the rays). 

As an explicit example (see also \citealt{Dessart05}), 
for a 2D grid of $n_{\rm r}$ radial and $n_\phi$ azimuthal 
points, one needs $n_{\rm r} n_\phi$ integrations of order $n_{\rm r} n_{\rm x}$ operations 
for every considered ray; this gives an overall scaling $n_{\rm ray} n_{\rm x} n_{\rm r}^2 n_\phi$, 
implying for a typical case of 
$n_{\rm ray} \approx 5$, $n_\phi \approx 100$, 
and $n_{\rm x} \approx n_{\rm r} \approx 1000$ on order $10^{11-12}$ 
operations to evaluate the radiative force. Moreover, such a 
calculation has to be 
carried out at \textit{each time-step of the hydrodynamical simulation}, 
which for a typical courant time $\sim 5$\,sec in a hot-star wind 
outflow, and a total simulation-time of, say, $\sim 50$ dynamical 
time scales $t_{\rm dyn} = R_\ast/\varv_\infty \sim10$\,ksec,  
requires some $\sim10^5$ repeated evaluations of the radiative force. 
This simple example thus illustrates quite vividly the rather daunting 
task of constructing multi-dimensional LDI wind models.        

Nonetheless, a few previous attempts have been performed. \citet{Dessart03} 
carried out `2D hydro+1D radiation' simulations by simply focusing only on the 
line force from a single radial ray, thus ignoring lateral influences. This led then 
to extensive break-up of the spherical shells seen in 1D simulations, 
resulting in lateral incoherence all the way down to the grid-scale. However, 
these simulations ignore the lateral component of the diffuse radiative 
force, which linear stability analysis \citep{Rybicki90} shows could lead 
to damping of velocity variations at scales below the lateral Sobolev 
length $\ell_{\rm Sob} = r\varv_{\rm th}/\varv$ and as such to more 
lateral coherence than seen in the single-ray 2D simulations. \citet{Dessart05} made 
a first attempt to include oblique rays, by using a special, restricted 
numerical grid in a 2D plane that forced 3 rays to always intersect 
the discrete mesh points \citep{Owocki99b}. But while these simulations did seem to suggest a 
somewhat larger lateral coherence than comparable 1-ray models, 
the inherent limitations of the method (e.g. in resolving the proper 
lateral scales) left results uncertain \citep{Dessart05}. 

This paper introduces a `pseudo-planar' modeling approach for a 
multi-dimensional wind subject to the LDI. In this 
`box-in-a-wind' method, all sphericity effects of 
the expanding flow are included in a radial direction $r$, but 
some curvature terms are ignored in the lateral direction(s). As discussed in 
\S\ref{modeling} (and detailed in Appendix A), for a 2D simulation in the $r,y$ plane this allows us to 
consider 5 `long characteristic' rays with a computational cost-scaling 
$3 n_{\rm x} n_{\rm r} n_{\rm y}$, thus reducing the general scaling above 
with a factor $\sim n_{\rm r} = 1000$ for our standard set-up. Using this method, 
\S\ref{results} examines the resulting 2D clumpy wind structure in 
much greater detail than possible before, and \S\ref{discussion} discusses the 
results, compares to other simulation test-runs, and outlines future work.    

\section{Modeling} 
\label{modeling} 

The simulations here use the numerical PPM  
\citep{Colella84} hydrodynamics code {\sc VH-1}\footnote{The 
{\sc VH-1} hydrodynamics computer-code package has been developed 
by J. Blondin and collaborators, and is available for download at:
http://wonka.physics.ncsu.edu/pub/VH-1/} 
to evolve the conservation equations of mass and 
momentum for a 2D, isothermal line-driven stellar wind outflow. A key point of this paper
is that while we keep all sphericity effects of an expanding outflow in the radial direction, we 
neglect some curvature terms in the lateral direction(s); for details, see Appendix A. Preserving 
all properties 
of a spherical outflow, this pseudo-planar, box-in-a-wind approach allows us to resolve 
laterally the relevant clump-length-scales, as well as implement non-radial rays for the radiative 
line-driving in a time-efficient way (see further below and Appendix A). 

All presented results adopt the same stellar and wind parameters as in \citet{Sundqvist13, Sundqvist15}, 
given here in Table 1, which are typical for an O-star in the Galaxy.
The standard set-up uses a spatial grid with 1000 discrete radial ($r$) 
mesh-points between $R_\ast \le r \le 2R_\ast$ and 100 lateral ($y$) 
ones that cover in total $0.1 R_\ast$. As such, the grid is uniform and 
has a constant step-size $\Delta = 0.001 R_\ast$; a small 
$\Delta$ is required to resolve both the sub-sonic wind-base with 
effective scale height $H = a^2R_\ast^2/(GM_\ast(1-\Gamma_{\rm e})) \approx 0.002 R_\ast > \Delta$ 
and the resulting small-scale 2D clump structures in the supersonic wind 
(the focus of this paper). Each simulation evolves from a smooth, CAK-like initial condition, 
computed by relaxing to a steady state a 1D spherically symmetric 
time-dependent simulation that uses a CAK/Sobolev form for the line-force.
To prevent artificial structure due to numerical truncation errors
we use an evolution time-step that is the minimum of a fixed 2.5~sec and 
a variable 1/3 of the courant time (see discussion in \citealt{Poe90}). 
As in previous work, the lower boundary at the assumed stellar 
surface fixes the density to a value $\sim 5-10$ times that at the 
sonic point. Moreover, since we are interested in structures that are 
considerably smaller than the computational box, the lateral boundaries 
are simply treated as periodic. 

\begin{table}
	\centering
	\caption{Summary of stellar and wind parameters}
		\begin{tabular}{p{2.8cm}ll}
		\hline \hline Name & Parameter & Value \\ 
                \hline
                Stellar luminosity & $L_\ast$ & 
                $8\,\times\,10^{5}\,\rm L_\odot$ \\ 
                Stellar mass & $M_\ast$ & 50 \,$\rm M_\odot$ \\ 
                Stellar radius & $R_\ast$ & 20 \,$\rm R_\odot$ \\
                Isoth. sound speed & $a$ & 23.4\,km/s \\ 
                Average & & \\
                \ - wind speed at $2R_\ast$& $\langle \varv_{\rm max} \rangle$ & 1230\,km/s \\ 
                \ - mass-loss rate & $\langle \dot{M} \rangle$ & 
                $1.3\,\times\,10^{-6}\,\rm M_\odot/yr$ \\ 
                CAK exponent & $\alpha$ & 0.65 \\ 
                Line-strength & & \\
                \ - normalization & $\bar{Q}$ & 2000 \\
                \ - cut-off & $Q_{\rm max}$ & 0.004$\bar{Q}$ \\               
                Ratio of ion thermal & & \\
                speed to sound speed & $\varv_{\rm th}/a$ & 0.28 \\ 
                Eddington factor & $\Gamma_{\rm e} = $ & 0.42 \\ 
                   & $\kappa_{\rm e} L_\ast/(4 \pi G M_\ast c)$ & \\
                \\
                \hline
		\end{tabular}
	\label{Tab:params}
\end{table}

\subsection{Radiative driving} 

\begin{figure}
      \resizebox{\hsize}{!}  {\includegraphics[]{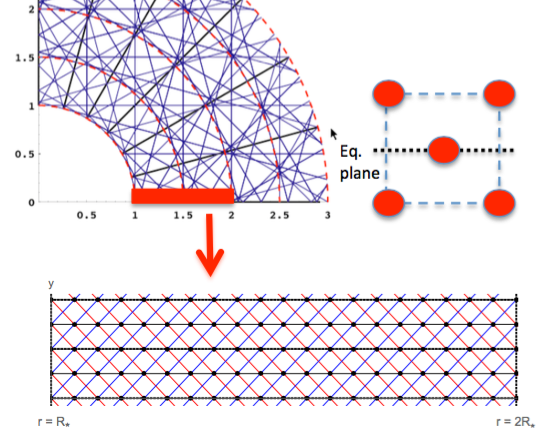}}
    \centering
  \caption{Sketch illustrating the basic idea of the pseudo-planar, box-in-a-wind approach 
  used in this paper. The upper left illustrates the general situation of non-alignment 
  between oblique rays and the numerical grid points. The lower panel then shows 
  how we create a pseudo-planar box in the wind by cutting out a small, but representative, 
  fraction of the wind volume. For illustration purposes, 
   it shows projections onto the equatorial plane of rays in the prograde (blue), retrograde (red), and radial (black) directions, for two lateral periods of a simple case with just $n_{\rm y}=2$ zones in lateral direction $y$. The right panel then illustrates how extension out of the equatorial plane involves a total of 5 rays: one radial plus two oblique pairs that extend up/down from the plane. 
   See Appendix A for a detailed explanation and for further illustrations of the assumed ray geometry.}
  \label{Fig:pp}
\end{figure}

The central challenge in these simulations is to compute the
2D radiation line-force in a highly structured, time-dependent wind with a
non-monotonic velocity. This requires non-local integrations of the
line-transport within each time-step of the simulation, in order to
capture the instability near and below the Sobolev length. To meet 
this objective, we develop here a multi-dimensional pseudo-planar 
extension of the smooth source function \citep[SSF,][]{Owocki91b} method 
described extensively in \citet{Owocki96} (see also \citealt{Sundqvist13}). 
Appendix A describes in detail this 2D-SSF formulation; below 
follows a summary of key features.   

Our pseudo-planar 2D-SSF approach allows us to follow the non-linear 
evolution of the strong, intrinsic LDI in the radial
direction, while simultaneously accounting for the potentially stabilizing effect of the 
scattered, diffuse radiation field, in \textit{both} the radial and lateral directions
\citep{Lucy84, Owocki85, Rybicki90}. SSF further assumes 
the line-strength number distribution to be given by an exponentially truncated power-law. 
In this formalism, $\alpha$ is the standard CAK power-law index, which can be physically 
interpreted as the ratio of the line force due to optically thick lines to the total line
force; $\bar{Q}$ is a line-strength normalization constant, which
can be interpreted as the ratio of the total line force to the
electron scattering force in the case that all lines were optically
thin; $Q_{\rm max}$ is the maximum line-strength cut-off\footnote{Note that we have recast the line force using the
  $\bar{Q}$ notation of \citet{Gayley00} rather than the $\kappa_0$
  notation of OP96. $\bar{Q}$ has the advantage of being a
  dimensionless measure of line-strength that is independent of the
  thermal speed. The relation between the two parameter formulations
  is given in Appendix A.}.
For typical O-star conditions at solar
metallicity, $Q_{\rm max} \approx \bar{Q} \approx 2000$
\citep{Gayley95, Puls00}. In practice, keeping the nonlinear amplitude
of the instability from exceeding the limitations of the numerical
scheme requires a significantly smaller cut-off \citep{Owocki88, Sundqvist13}.
  
As noted
in the introduction, including oblique rays 
in a multi-dimensional outflow presents severe computational 
challenges, largely due to the general misalignment of the 
rays with the nodes of the numerical grid. While 
earlier attempts of 2D LDI simulations have either used a 
`2D-hydro 1D-radiation' approach \citep{Dessart03} or experimented 
with a restricted special radial grid set-up \citep{Dessart05}, the pseudo-planar 
method introduced here largely circumvents these 
issues of grid-misalignment. Namely, while radial ray-integrations are 
here calculated identically to the original SSF method, for oblique rays
both the azimuthal radiation angle $\phi$ and the ray's radial directional cosine 
$\mu \equiv \cos \theta = \hat{r} \cdot \hat{n}$ become 
constant throughout the computational domain. 
To this end, we apply a set of 5 rays with $\mu, \phi = 
(1,1/\sqrt{3},1/\sqrt{3},1/\sqrt{3},1/\sqrt{3})\,,\,(0,\pi/4,-\pi/4,3\pi/4,-3\pi/4)$
(see simple illustration in Fig. \ref{Fig:pp}, and Appendix A for a detailed explanation). 
In addition to the (trivial) radial ray, this thus 
considers 4 oblique rays that are also pointing up/down with respect to 
the 2D equatorial plane in which the hydrodynamical calculations are carried 
out (in order to avoid certain 2D `flat-land' radiation effects, see 
\citealt{Gayley00}). 
For our assumed grid then, with constant spacings in radial and lateral directions, 
information can be used for \textit{all} grid nodes when the 
ray-integration for a given $(\mu,\phi)$ pair has been 
performed only once over $R_\ast \le r \le 2R_\ast$ for each 
of the lateral grid-points. This means that the solid 
angle integrations required to compute the 
line-force in the radial and lateral directions then can be performed  
without the need of any further ray-integrations. In addition, because 
of the symmetry of rays pointing up/down from the equatorial plane, 
we only have to explicitly carry out the integrations for 3 of our 
5 angles. With respect to the general situation, this  
means we have effectively reduced the number of required 
`long characteristic' ray-integrations at each time-step with a factor of $\sim n_r$
($=10^3$ for our standard set-up here)!   

Another attractive feature of this pseudo-planar model is that it preserves 
all properties for a 1D purely radial outflow. As detailed in Appendix A, 
this is achieved by preserving the general scaling of the flux with radius for a spherical outflow, 
by including a sink term for the density to mimic spherical divergence, and by including 
in the force equations terms to account for stellar rotation along the lateral axis $y$. As such, our 
approach allows for easy testing and benchmarking, and we have 
verified that a simulation run with $n_y=1$ and 
integration weights for all oblique rays set to zero indeed gives the same results as a 
`normal' 1D spherical radial-ray SSF simulation. However, since such radial models 
are also subject to the global wind instability associated with nodal topology 
\citep{Poe90, Sundqvist15}, they exhibit clumpy structure all the 
way down to the lower boundary \citep{Sundqvist13, Sundqvist15}. While there 
are strong indications that clumping in hot star winds indeed 
extends to very near-photospheric layers \citep[e.g.,][]{Cohen14}, 
in these first 2D simulations we nonetheless opt to stabilize the wind base by
introducing a small radial increase in $\bar{Q}$ between $R_\ast < r < 1.5 R_\ast$.
This allows us study the emerging clump formation and structure in a somewhat 
more controlled environment as compared to simulations 
that lie on the nodal topology branch (see \S\ref{discussion}). 

\section{Simulation results}
\label{results} 

\begin{figure*}
    \vspace{1cm}
	\begin{minipage}{18.0cm}
      \resizebox{\hsize}{!}  {\includegraphics[]{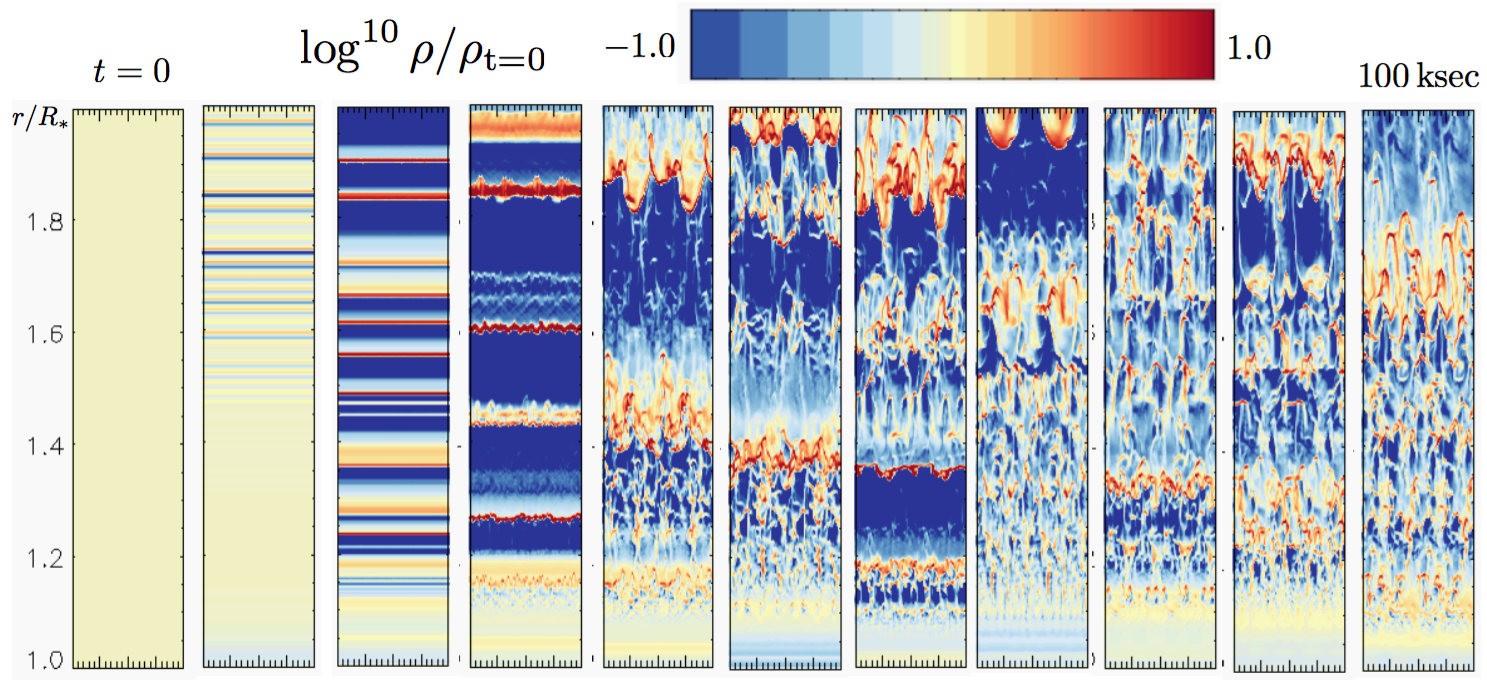}}
    \centering
    \end{minipage}
    \vspace{1cm}
    	\begin{minipage}{18.0cm}
      \resizebox{\hsize}{!}  {\includegraphics[]{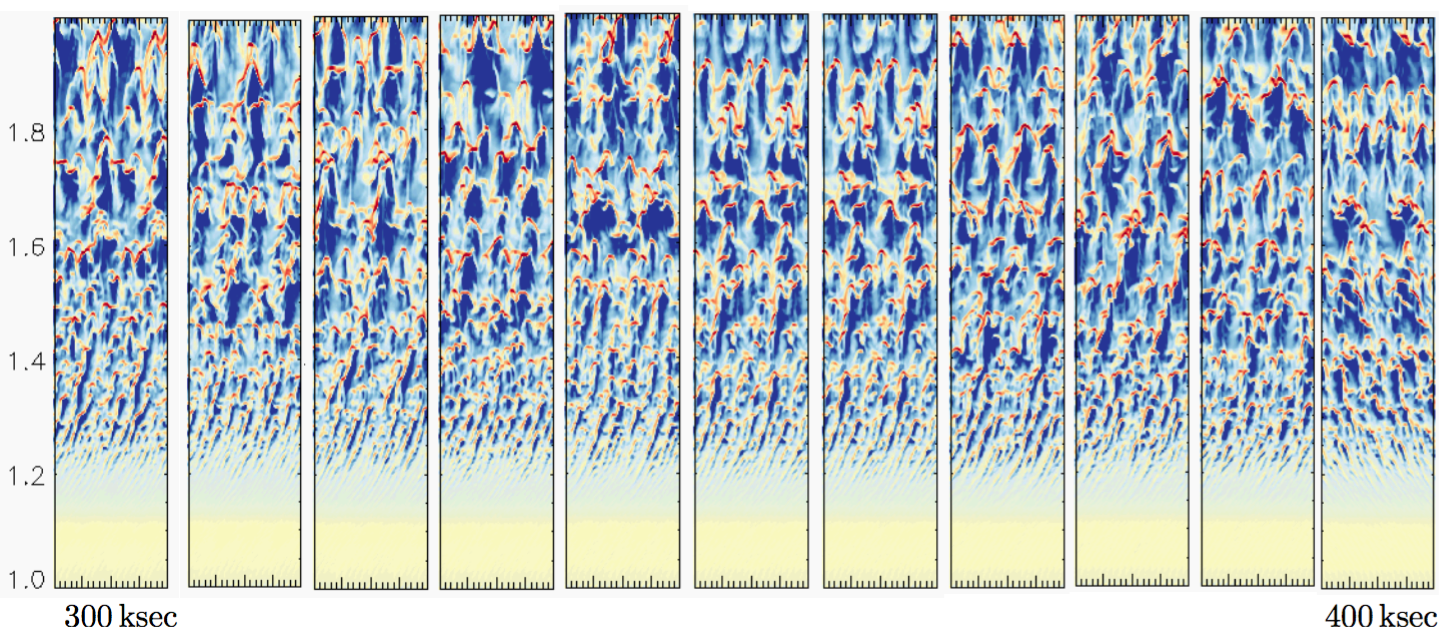}}
    \centering
    \end{minipage}
  \caption{Spatial and temporal variations of log density relative to the initial, smooth 
  `CAK' steady-state at $t=0$, with color ranging from densities a decade 
  below the $t=0$ value (blue) to a decade above (red). The vertical variation extends 
  from the subsonic wind-base at the stellar surface $R_\ast$ to a height of one $R_\ast$ 
  above. For clarity, the lateral variation is displayed over \textit{twice} the horizontal box 
  length $0.1 R_\ast$. The upper row shows time evolution over the initial 100 ksec after the 
  CAK initial condition, in steps of 10 ksec; the bottom row uses the same step-size of 10 ksec 
  to show the evolution between 300 and 400 ksec, long after the initial condition has developed 
  into a statistically steady turbulent flow.}       
  \label{Fig:rho}
\end{figure*}

\begin{figure*}
    \vspace{1cm}
      \resizebox{\hsize}{!}  {\includegraphics[]{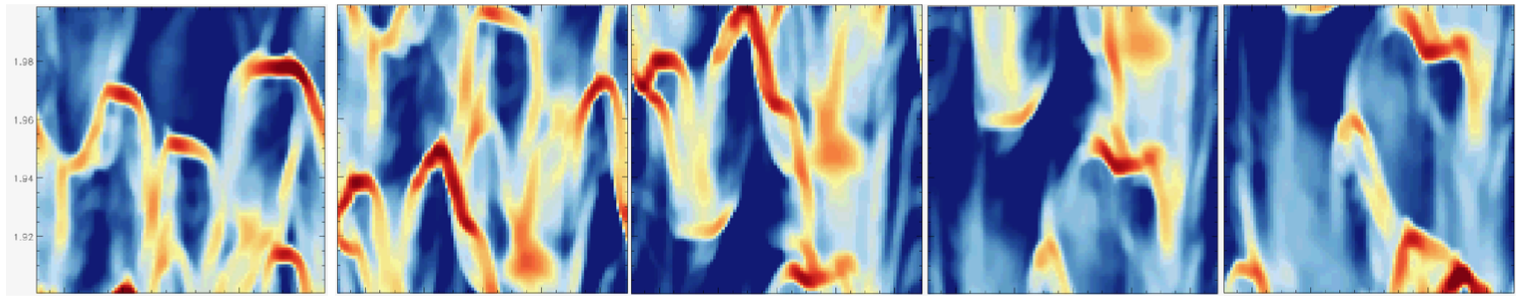}}
    \centering
  \caption{As in Fig. \ref{Fig:rho}, spatial and temporal variations of log density relative to the initial, smooth 
  `CAK' steady-state at $t=0$ are shown, with color ranging from densities a decade 
  below the $t=0$ value (blue) to a decade above (red). Here the vertical variation only extends between $1.9R_\ast$ and
  $2.0R_\ast$ and the lateral variation is displayed over \textit{one} horizontal box of $0.1 R_\ast$; there are thus $100 \times 100$ discrete mesh-points in each of the displayed squares. From left to right are shown a 2 ksec time-evolution long after the initial 
  condition, in steps of 0.5 ksec.}  
  \label{Fig:rho_zoom}
\end{figure*}

\begin{figure*}
    \vspace{1cm}
      \resizebox{\hsize}{!}  {\includegraphics[]{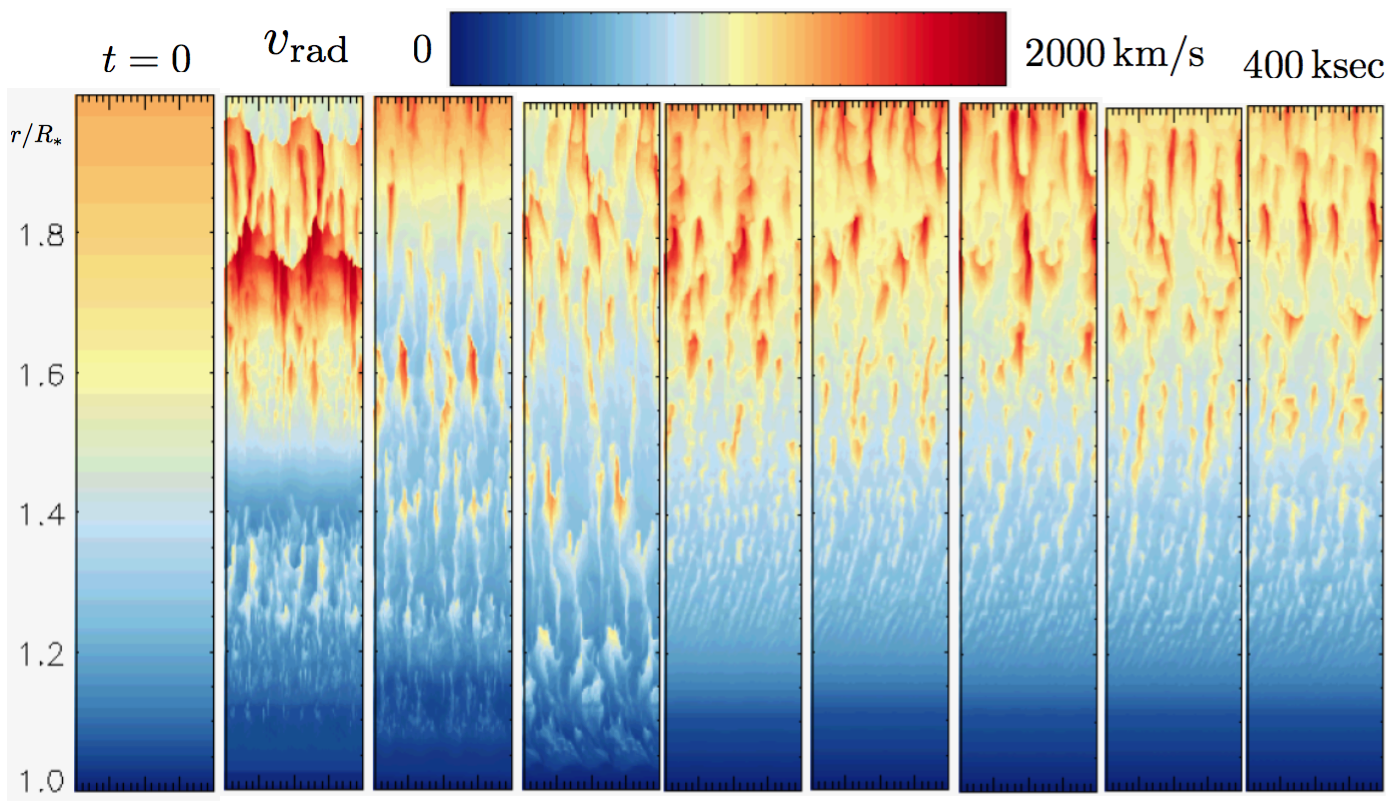}}
    \centering
  \caption{Spatial and temporal variations of radial velocity $\varv_{\rm rad}$, with color ranging 
  from 0 (blue) to 2000\,$\rm km/s$ (red). As in Fig. \ref{Fig:rho}, the vertical variation extends 
  from the subsonic wind-base at the stellar surface $R_\ast$ to a height of one $R_\ast$ 
  above, and the lateral variation is displayed over \textit{twice} the horizontal box 
  length $0.1 R_\ast$. The frames from left to right show the time evolution of $v_{\rm rad}$ over
  400 ksec after the CAK initial condition, in steps of 50 ksec.}   
  \label{Fig:vrad}
\end{figure*} 

\begin{figure*}
    \vspace{1cm}
    \resizebox{\hsize}{!}  {\includegraphics[]{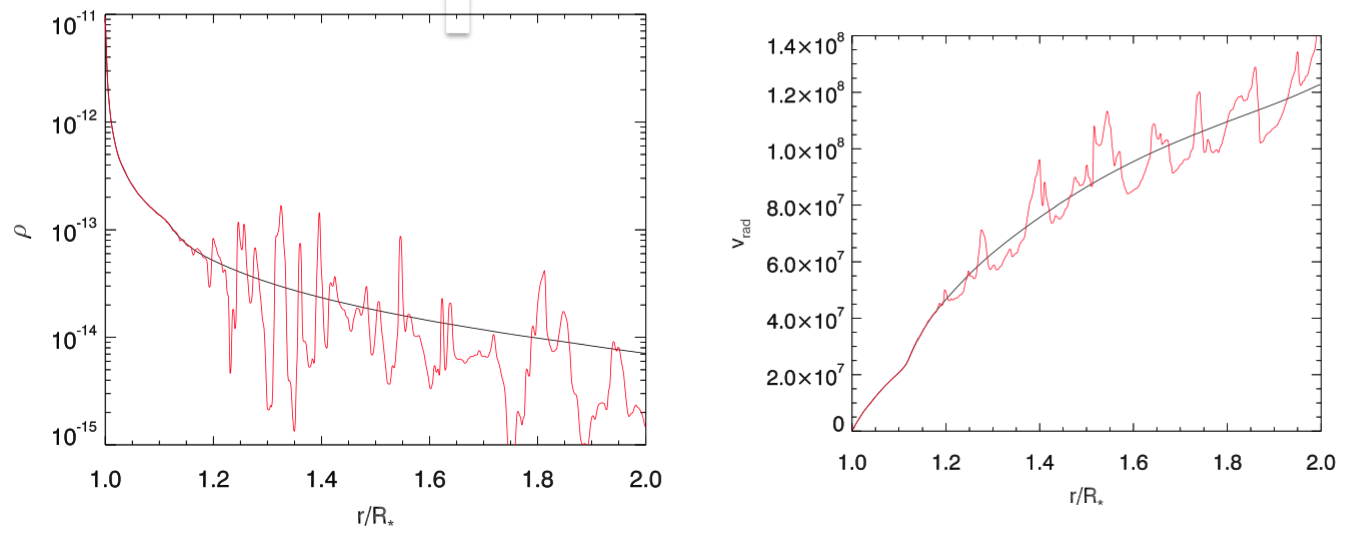}}
    \centering
  \caption{Radial cuts through the 2D simulation box of density 
  $\rho \, \rm [g/cm^3]$ (left) and radial velocity $\varv_{\rm rad} \, \rm [cm/s]$ (right). The red curves 
  are taken at a snapshot long after the simulation has developed into a statistically quite steady flow; the black 
  curves compare this to average values.}   
  \label{Fig:1d-cut}
\end{figure*}

Fig. \ref{Fig:rho} illustrates directly a key result of our simulations, namely 
the spatial and temporal variation in $\log \rho$ 
relative to the initial, smooth `CAK' steady-state. The figure 
shows clearly how a radial shell structure first develops, but then quickly 
breaks up into laterally complex density variations. The upper 
panel displays snapshots during the first 100 
$\rm ksec$ of the simulation, illustrating how
already after a few dynamical flow-times $t_{\rm dyn} \approx R_\ast/\langle \varv_{\rm max} \rangle 
\approx \rm 11 \, ksec$ the characteristic shells, seen in all 
1D LDI simulations, brake up in what initially seem to resemble 
Rayleigh-Taylor structures. The lower panel then shows how, as 
time passes by, the structures eventually develop into a complex but 
statistically quite steady flow, characterized now by localized density 
enhancements (`clumps') of very small spatial 
scales embedded in larger regions of much lower density. 

Fig. \ref{Fig:rho_zoom} zooms in on the same log density in a
small $0.1R_\ast$ square-box over a short time-sequence long 
after the initial condition. This illustrates in greater detail 
the quite complex 2D density structure, showing a range of 
scales as well as high-density clumps with different 
shapes. The figure also demonstrates that, 
although the structures are small, they are clearly resolved by 
our numerical grid. 

Fig. \ref{Fig:vrad} displays 
temporal and spatial variations in radial velocity, 
illustrating essentially the same kind of
outer-wind shock structure and high velocity streams as 
corresponding 1D simulations; however, also the velocity 
now exhibits extensive lateral variation, reflecting again 
the break-up of 1D shells into small-scale 2D clumps. 

Fig. \ref{Fig:1d-cut} emphasizes some similarities between these 
2D simulations and corresponding 1D ones, by  
showing a radial cut through the simulation box at 
a time-snapshot (again taken long after the simulation has developed 
into a statistically steady flow). The figure demonstrates 
how such radial cuts indeed still show the characteristic 
structure of the non-linear growth of the LDI, namely 
high-speed rarefactions that steepen into strong shocks
and wind plasma compressed into spatially narrow `clumps' 
separated by rather large regions of rarified gas. There are some 
differences though: In addition to the lateral break-up of shells discussed above, another key 
distinction between 1D and 2D simulations is that the radial density variations are a 
bit lower in the latter; this occurs because of the lateral `filling in' of radial 
rarefactions \citep[see also][]{Dessart03} and is discussed further in the following 
section.  

\subsection{Statistical properties} 

\begin{figure*}
    \vspace{1cm}
    \resizebox{\hsize}{!}  {\includegraphics[]{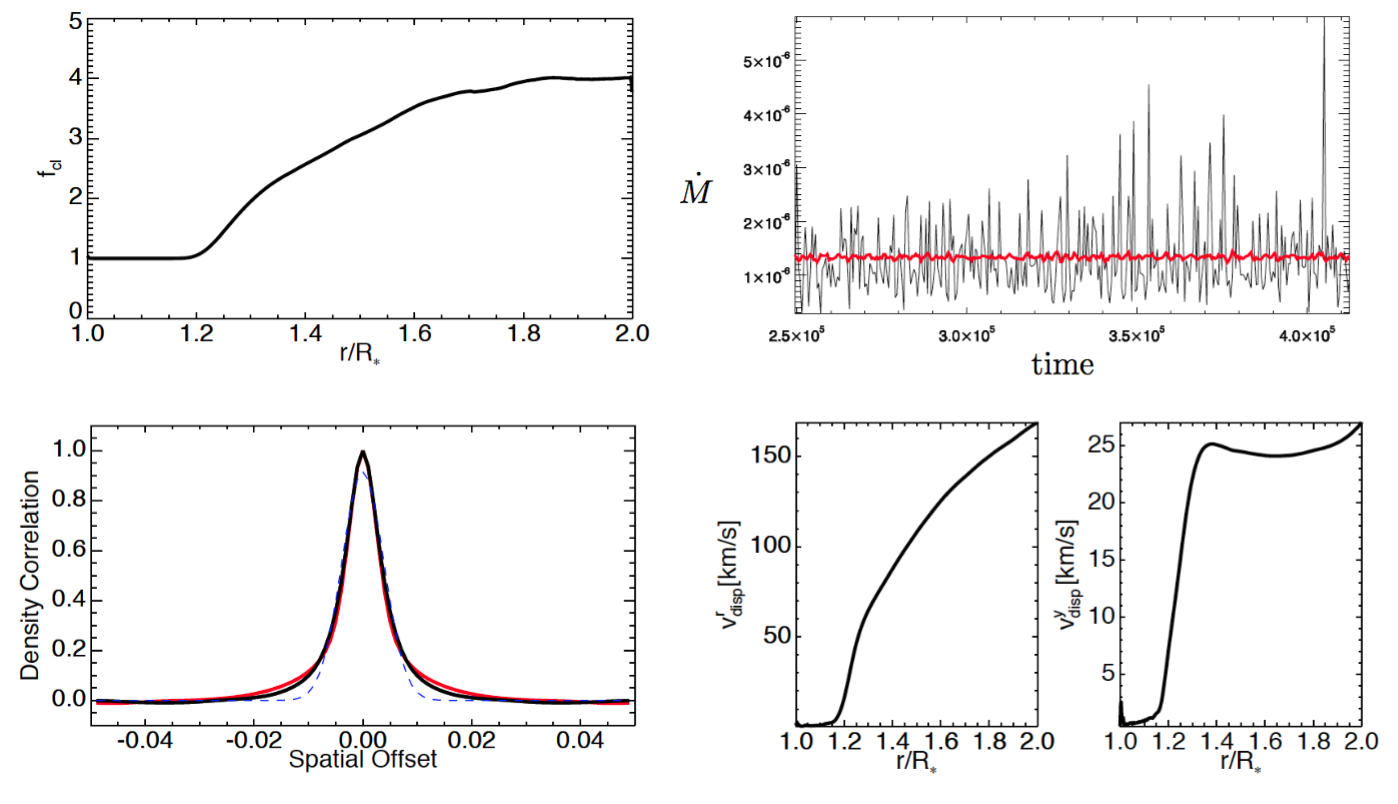}}
    \centering
  \caption{Selected statistical properties of the 2D simulation, see text. The upper left panel plots 
  the clumping factor $f_{\rm cl}$; the upper right panel shows the time-dependent 
  mass-loss rate, $\dot{M} \, \rm [M_\odot/yr]$ vs. sec., computed in two different ways for the red and black curves (see text); 
  the lower left panel displays lateral (black) and radial (red) density correlation lengths as well as a Gaussian fit to these (blue, dashed); the lower right panel then finally plots radial (left) and lateral (right) velocity dispersions.}   
  \label{Fig:stat}
\end{figure*}

Fig. \ref{Fig:stat} summarizes some statistical results of the simulations. All averaging 
have here started at $t = 250 \, \rm ksec$, in order to separate out any dependence 
on the initial conditions and the adjustment to a new radiative 
force balance. The upper left panel in Fig. \ref{Fig:stat} shows the clumping factor: 
\begin{equation} 
	f_{\rm cl} = \frac{\langle \rho^2 \rangle}{\langle \rho \rangle^2},  
\end{equation} 
where angle brackets denote averaging both laterally and 
over time in order to separate out $f_{\rm cl}$'s primary dependence on radius. 
The plot illustrates how the lateral `filling-in' of radially compressed gas (see above)
decreases the quantitative clumping factor significantly in a 2D simulation as compared to earlier 1D 
models where $f_{\rm cl} \ga 10$ \citep[e.g.,][]{Sundqvist13}; 
this is also consistent with the previous 2D results by \citet{Dessart03}. Note, however, 
that the actual values of $f_{\rm cl}$ in our 2D simulation are likely somewhat 
underestimated, due to our choice of stabilizing the wind-base against instability caused by
nodal topology (see previous section). As 
discussed extensively by \citet{Sundqvist13}, in these near-photospheric layers the 
quantitative clumping factor is very sensitive to such choices made for the calculation 
of the radiative acceleration, as well as to any variability that may be assumed for the 
photospheric lower boundary. Regardless of such caveats, the basic qualitative result 
here that 2D simulations yield relatively lower values of $f_{\rm cl}$ than 
comparable 1D simulations is quite robust. 

The upper right panel of Fig. \ref{Fig:stat} then shows the time-dependent mass-loss rate:  
\begin{equation} 
	\dot{M} \equiv 4 \pi r^2 \rho \varv_{\rm rad}.
\end{equation} 
The black line in this plot shows a simple lateral average of the mass-flux escaping the 
outermost radial grid-point at a specific time. However, since our simulation box 
only covers $0.1 R_\ast$, such an average very likely overestimates the time-dependent 
mass loss significantly. To compensate for this, the red curve in the plot instead 
uses an average over all grid-points $r \ge 1.5 R_\ast$ at a specific time, 
which approximates averaging over a full stellar surface $4 \pi (2 R_\ast)^2 \approx 50 R_\ast^2.$ 
As expected, this curve shows a drastically lower temporal variation of $\dot{M}$, despite the 
highly time-dependent flow. This is consistent e.g. with decade-long observations of spectral lines in 
O-stars, which typically indicate that time-variations in the mass-loss rate of such stars are low.   

To estimate typical clump length-scales, the lower left panel of Fig. \ref{Fig:stat} 
plots a density autocorrelation length: 
\begin{equation} 
	f_{\rm c}(\Delta) = \sum_{\rm time} \sum_{\rm i} 
	(\rho_i - \langle \rho \rangle) \, (\rho_{i-\Delta} - \langle \rho \rangle), 
\end{equation} 
where $\langle \rho \rangle$ averages laterally and over time. A lateral correlation
length is calculated at each of the $\Delta = 0-99$ lateral mesh-points 
and normalized to its $\Delta = 0$ value. The figure then plots an 
average of this lateral correlation length between $r/R_\ast = 1.9-2.0$ 
(black curve), as well as a radial correlation length (red curve)
defined analogously. The lateral and radial density correlation lengths are very similar, 
and as such illustrates how a statistical ensemble of clumps is quite isotropic in 
these simulations. This does not imply that 
any given clump is isotropic (see Fig. \ref{Fig:rho_zoom}), but rather that, on 
average, the well-developed density variations in the simulations do not have a 
strong preferred direction. 

The Gaussian fit plotted in the blue dashed curve provides an estimate of the 
autocorrelation length in terms of the gaussian FWHM $\approx 0.01 R_\ast$. 
Such small characteristic scales agree well with the theoretical expectation (see introduction) 
that the critical length scale for these clumpy wind simulations is of order the 
Sobolev length $\ell_{\rm Sob}$, which for the lateral direction at $2  R_\ast$ is 
$\ell_{\rm Sob}/R_\ast = 2\varv_{\rm th}/\varv \approx 0.01$. 
Identifying this as a typical clump length scale $\ell_{\rm cl}$, we may further make a 
simple estimate of the typical clump mass $\ell_{\rm cl}^3 \rho_{\rm cl} \approx 
10^{-6} R_\ast^3 \,  7 \times 10^{-14} \, \rm g/cm^3 \approx 10^{17} \, \rm g$, where 
the estimated clump density here simply reads off the output of the simulations 
(e.g., Fig. \ref{Fig:1d-cut}). More generally, such a clump 
mass-estimate may be obtained using the Sobolev length and mass conservation: 
\begin{equation} 
m_{\rm cl} \approx \ell_{\rm Sob}^3 \rho_{\rm cl} 
\approx \frac{\varv_{\rm th}^3 \dot{M} f_{\rm cl} r}{\varv^4 4 \pi}, 
\label{Eq:mcl}
\end{equation}    
which for the 2D simulation analyzed here indeed 
gives $m_{\rm cl} \approx 10^{17} \, \rm g$ for typical 
values at $2 R_\ast$. Quite generally, eqn. \ref{Eq:mcl} shows
explicitly how rather low clump masses are expected to emerge 
from the LDI.    
  
Finally, the lower right panels in Fig. \ref{Fig:stat} plots the radial and 
lateral velocity dispersions: 
\begin{equation} 
	\varv_{\rm disp} = \sqrt{\langle \varv^2 \rangle - \langle \varv \rangle^2}, 
\end{equation}
where averages are constructed like for the clumping 
factor above. These plots show how, as expected (see also \citealt{Dessart03}), 
the lateral velocity dispersion is on order the isothermal sound speed, 
whereas the radial dispersion is much higher and 
expected to rise above several hundreds $ \rm km/s$ in the outer wind. 

\section{Summary and future work} 
\label{discussion}

We have introduced a pseudo-planar, box-in-a-wind approach suitable for
carrying out radiation-hydrodynamical simulations in situations where the 
computation of the radiative acceleration is 
challenging and time-consuming. The method is used here to 
simulate the 2D non-linear evolution of the strong  
line-deshadowing instability (LDI) that causes clumping in the 
stellar winds from hot, massive stars. Accounting fully for both 
the direct and diffuse radiation components in the calculations of 
both the radial and lateral radiative accelerations, we examine in 
detail the small-scale clumpy wind structure resulting from our simulations. 

Overall, the 2D simulations show that the LDI first manifests itself
by mimicking the typical shell-structure seen in 1-D simulations, but 
these shells then quickly break up because of 
basic hydrodynamic instabilities like Rayleigh-Taylor and influence 
of the oblique radiation rays. This results in a quite 
complex 2D density and velocity structure, characterized by 
small-scale density `clumps' embedded in larger regions of fast and 
rarefied gas. 

While inspection of radial cuts through the 2D simulation box confirms 
that the typical radial structure of the LDI is intact, quantitatively the 
lateral `filling-in' of gas leads to lower values of the clumping factor 
than for corresponding 1D models. A correlation-length 
analysis further shows that, statistically, density-variations in the well-developed 
wind are quite isotropic; identifying then the computed autocorrelation length with a 
typical clump size gives $\ell_{\rm cl}/R_\ast \sim 0.01$ at $2 R_\ast$, 
and thus also quite low typical clump-masses $m_{\rm cl} \sim 10^{17}$\,g.
This agrees well with the theoretical expectation that the important 
length-scale for LDI-generated wind-structure is of order the Sobolev 
length $\ell_{\rm Sob}$. 

\begin{figure*}
    \vspace{1cm}
      \resizebox{\hsize}{!}  {\includegraphics[]{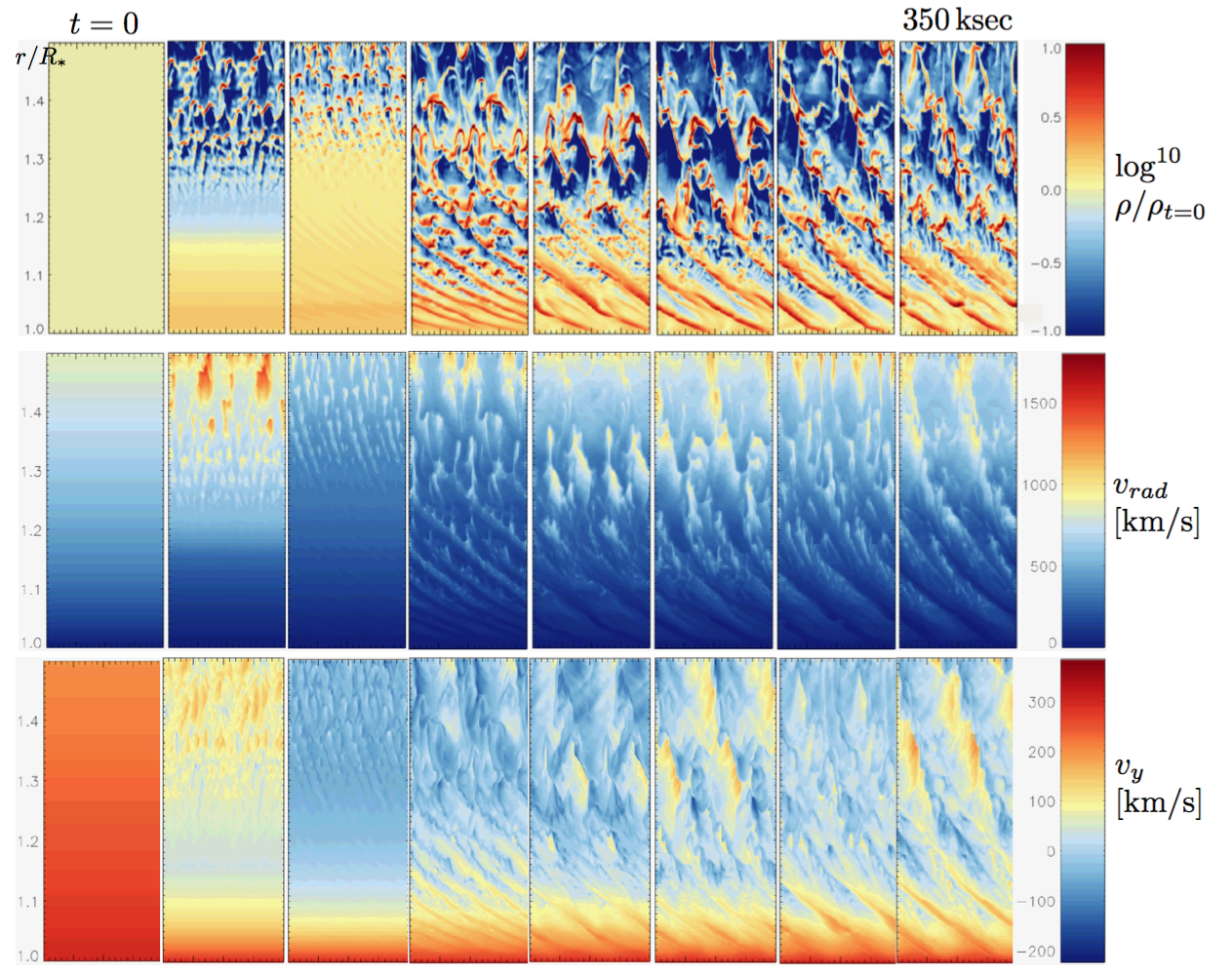}}
    \centering
  \caption{Spatial and temporal variations of log density, radial velocity, and lateral velocity for a 
  model with stellar rotation at the surface $\varv_{\rm y} = 300$\,km/s (see text), with color ranging 
  as in the earlier Figs. \ref{Fig:rho} and \ref{Fig:vrad}. The vertical variation in this simulation 
  extends only from 1.0-1.5 $R_\ast$, but the lateral variation is displayed as before 
  over \textit{twice} the horizontal box length $0.1 R_\ast$. From left to right are shown the time evolution 
  over 350 ksec after the CAK initial condition, in steps of 50 ksec.}   
  \label{Fig:rotation}
\end{figure*} 

\paragraph{Influence of rotation and topology.} As noted in 
\S2 and \S3.1, the level of structure in near photospheric 
layers is likely underestimated in the simulation analyzed above, 
due to our choice to stabilize the wind base.  
To demonstrate this further, Fig. \ref{Fig:rotation} shows a test-run with 
identical 2D set-up as before, but now introducing stellar rotation with 
a fixed $\varv_{\rm rot} = 300$\,km/s at the surface,  
and an initial condition set by steady-state angular momentum 
conservation, $\varv_{\rm y}(r) = \varv_{\rm rot} R_\ast/r$. The 
figure shows that once the simulation has adjusted to its new 
force conditions, radial streaks of high density now appear already 
at the surface; in other test-runs, we have found that such 
structures are typical for simulations with an unstable base 
and nodal topology. The radial streaks in this 
rotating model migrate along with the surface rotation, and embedded in 
the larger-scale structures are the typical small-scale clumps 
discussed previously. As speculated already in 
\citet{Sundqvist15}, these tentative first results thus suggest 
that rotating LDI models may quite naturally lead to 
the type of combined large- and small-scale 
structure needed to explain in parallel various observed phenomena
in hot-star winds, like discrete absorption components (DACs) 
\citep{Kaper99} and small-scale wind clumping \citep{Eversberg98}. Future work will examine 
in detail connections between these rotating LDI models and the 
presence of various types of wind sub-structure.

The simulations presented in this paper also lead naturally to a number of follow-up 
investigations; already in the pipe-line are the development of a formalism for characterizing 
porosity-effects in turbulent media (Owocki \& Sundqvist 2017) and the influence of 
the clumpy wind on the accretion properties of an orbiting neutron star in a 
so-called high-mass X-ray binary (HMXB) system (el-Mellah et al. 2017). More 
directly related to this paper, we also plan to (in addition to further analyzing the effects of rotation and topology) 
extend the current simulations to 3D and to higher wind radii, and also develop a 
more general radiative transfer scheme (allowing for an arbitrary number of rays) for 
the computation of the line acceleration within a pseudo-planar box-in-a-wind.

\begin{acknowledgements}
  This work was supported in part by 
  SAO Chandra grant TM3-14001A awarded to the 
  University of Delaware, and in part by the visiting 
  professor scholarship ZKD1332-00-D01 for SPO 
  from KU Leuven. SPO acknowledges sabbatical leave 
  support from the University of Delaware, and we 
  also thank John Castor for helpful discussions on 
  long-characteristic methods. We finally thank the 
  referee for useful comments on the paper.    
\end{acknowledgements}

\bibliographystyle{aa}
\bibliography{2d_ldi}

\begin{thebibliography}{33}
\expandafter\ifx\csname natexlab\endcsname\relax\def\natexlab#1{#1}\fi

\bibitem[{{Berghoefer} {et~al.}(1997){Berghoefer}, {Schmitt}, {Danner}, \&
  {Cassinelli}}]{Berghofer97}
{Berghoefer}, T.~W., {Schmitt}, J.~H.~M.~M., {Danner}, R., \& {Cassinelli},
  J.~P. 1997, \aap, 322, 167

\bibitem[{{Castor} {et~al.}(1975){Castor}, {Abbott}, \& {Klein}}]{Castor75}
{Castor}, J.~I., {Abbott}, D.~C., \& {Klein}, R.~I. 1975, \apj, 195, 157

\bibitem[{{Cohen} {et~al.}(2010){Cohen}, {Leutenegger}, {Wollman},
  {Zsarg{\'o}}, {Hillier}, {Townsend}, \& {Owocki}}]{Cohen10}
{Cohen}, D.~H., {Leutenegger}, M.~A., {Wollman}, E.~E., {et~al.} 2010, \mnras,
  405, 2391

\bibitem[{{Cohen} {et~al.}(2014){Cohen}, {Wollman}, {Leutenegger}, {Sundqvist},
  {Fullerton}, {Zsarg{\'o}}, \& {Owocki}}]{Cohen14}
{Cohen}, D.~H., {Wollman}, E.~E., {Leutenegger}, M.~A., {et~al.} 2014, \mnras,
  439, 908

\bibitem[{{Colella} \& {Woodward}(1984)}]{Colella84}
{Colella}, P. \& {Woodward}, P.~R. 1984, Journal of Computational Physics, 54,
  174

\bibitem[{{Dessart} \& {Owocki}(2003)}]{Dessart03}
{Dessart}, L. \& {Owocki}, S.~P. 2003, \aap, 406, L1

\bibitem[{{Dessart} \& {Owocki}(2005{\natexlab{a}})}]{Dessart05}
{Dessart}, L. \& {Owocki}, S.~P. 2005{\natexlab{a}}, \aap, 437, 657

\bibitem[{{Dessart} \& {Owocki}(2005{\natexlab{b}})}]{Dessart05b}
{Dessart}, L. \& {Owocki}, S.~P. 2005{\natexlab{b}}, \aap, 432, 281

\bibitem[{{Eversberg} {et~al.}(1998){Eversberg}, {Lepine}, \&
  {Moffat}}]{Eversberg98}
{Eversberg}, T., {Lepine}, S., \& {Moffat}, A.~F.~J. 1998, \apj, 494, 799

\bibitem[{{Feldmeier} {et~al.}(1997){Feldmeier}, {Puls}, \&
  {Pauldrach}}]{Feldmeier97}
{Feldmeier}, A., {Puls}, J., \& {Pauldrach}, A.~W.~A. 1997, \aap, 322, 878

\bibitem[{{Gayley}(1995)}]{Gayley95}
{Gayley}, K.~G. 1995, \apj, 454, 410

\bibitem[{{Gayley} \& {Owocki}(2000)}]{Gayley00}
{Gayley}, K.~G. \& {Owocki}, S.~P. 2000, \apj, 537, 461

\bibitem[{{G{\"u}del} \& {Naz{\'e}}(2009)}]{Gudel09}
{G{\"u}del}, M. \& {Naz{\'e}}, Y. 2009, \aapr, 17, 309

\bibitem[{{Kaper} {et~al.}(1999){Kaper}, {Henrichs}, {Nichols}, \&
  {Telting}}]{Kaper99}
{Kaper}, L., {Henrichs}, H.~F., {Nichols}, J.~S., \& {Telting}, J.~H. 1999,
  \aap, 344, 231

\bibitem[{{L{\'e}pine} \& {Moffat}(2008)}]{Lepine08}
{L{\'e}pine}, S. \& {Moffat}, A.~F.~J. 2008, \aj, 136, 548

\bibitem[{{Lucy}(1983)}]{Lucy83}
{Lucy}, L.~B. 1983, \apj, 274, 372

\bibitem[{{Lucy}(1984)}]{Lucy84}
{Lucy}, L.~B. 1984, \apj, 284, 351

\bibitem[{{Mart{\'{\i}}nez-N{\'u}{\~n}ez}
  {et~al.}(2017){Mart{\'{\i}}nez-N{\'u}{\~n}ez}, {Kretschmar}, {Bozzo},
  {Oskinova}, {Puls}, {Sidoli}, {Sundqvist}, {Blay}, {Falanga}, {F{\"u}rst},
  {G{\'{\i}}menez-Garc{\'{\i}}a}, {Kreykenbohm}, {K{\"u}hnel}, {Sander},
  {Torrej{\'o}n}, \& {Wilms}}]{Martinez17}
{Mart{\'{\i}}nez-N{\'u}{\~n}ez}, S., {Kretschmar}, P., {Bozzo}, E., {et~al.}
  2017, \ssr

\bibitem[{{Owocki}(1991)}]{Owocki91b}
{Owocki}, S.~P. 1991, in NATO ASIC Proc. 341: Stellar Atmospheres - Beyond
  Classical Models, ed. L.~{Crivellari}, I.~{Hubeny}, \& D.~G. {Hummer}, 235

\bibitem[{{Owocki}(1999)}]{Owocki99b}
{Owocki}, S.~P. 1999, in Lecture Notes in Physics, Berlin Springer Verlag, Vol.
  523, IAU Colloq. 169: Variable and Non-spherical Stellar Winds in Luminous
  Hot Stars, ed. B.~{Wolf}, O.~{Stahl}, \& A.~W. {Fullerton}, 294

\bibitem[{{Owocki} {et~al.}(1988){Owocki}, {Castor}, \& {Rybicki}}]{Owocki88}
{Owocki}, S.~P., {Castor}, J.~I., \& {Rybicki}, G.~B. 1988, \apj, 335, 914

\bibitem[{{Owocki} \& {Puls}(1996)}]{Owocki96}
{Owocki}, S.~P. \& {Puls}, J. 1996, \apj, 462, 894

\bibitem[{{Owocki} \& {Rybicki}(1984)}]{Owocki84}
{Owocki}, S.~P. \& {Rybicki}, G.~B. 1984, \apj, 284, 337

\bibitem[{{Owocki} \& {Rybicki}(1985)}]{Owocki85}
{Owocki}, S.~P. \& {Rybicki}, G.~B. 1985, \apj, 299, 265

\bibitem[{{Poe} {et~al.}(1990){Poe}, {Owocki}, \& {Castor}}]{Poe90}
{Poe}, C.~H., {Owocki}, S.~P., \& {Castor}, J.~I. 1990, \apj, 358, 199

\bibitem[{{Puls} {et~al.}(1993){Puls}, {Owocki}, \& {Fullerton}}]{Puls93}
{Puls}, J., {Owocki}, S.~P., \& {Fullerton}, A.~W. 1993, \aap, 279, 457

\bibitem[{{Puls} {et~al.}(2000){Puls}, {Springmann}, \& {Lennon}}]{Puls00}
{Puls}, J., {Springmann}, U., \& {Lennon}, M. 2000, \aaps, 141, 23

\bibitem[{{Puls} {et~al.}(2008){Puls}, {Vink}, \& {Najarro}}]{Puls08}
{Puls}, J., {Vink}, J.~S., \& {Najarro}, F. 2008, \aapr, 16, 209

\bibitem[{{Rybicki} {et~al.}(1990){Rybicki}, {Owocki}, \& {Castor}}]{Rybicki90}
{Rybicki}, G.~B., {Owocki}, S.~P., \& {Castor}, J.~I. 1990, \apj, 349, 274

\bibitem[{{Sobolev}(1960)}]{Sobolev60}
{Sobolev}, V.~V. 1960, {Moving envelopes of stars} (Cambridge: Harvard
  University Press, 1960)

\bibitem[{{Sundqvist} \& {Owocki}(2013)}]{Sundqvist13}
{Sundqvist}, J.~O. \& {Owocki}, S.~P. 2013, \mnras, 428, 1837

\bibitem[{{Sundqvist} \& {Owocki}(2015)}]{Sundqvist15}
{Sundqvist}, J.~O. \& {Owocki}, S.~P. 2015, \mnras, 453, 3428

\bibitem[{{Sundqvist} {et~al.}(2010){Sundqvist}, {Puls}, \&
  {Feldmeier}}]{Sundqvist10}
{Sundqvist}, J.~O., {Puls}, J., \& {Feldmeier}, A. 2010, \aap, 510, 11

\end{thebibliography}

\newpage
\newpage
\appendix 
\section{2D pseudo-planar line-force}
\label{sec:appa}

Our development here of a 2D vector form for the line-acceleration follows a direct generalization of the 1D SSF method detailed in Owocki and Puls (1996; hereafter OP96), as further developed in Sundqvist and Owocki (2015, hereafter SO15).
As discussed in OP96 (cf. their equation (3)), a key step is to compute efficiently the profile-weighted line optical depth between two wind locations along some ray coordinate $z$,
\beq
\Delta 
t(x,z_1,z_2) = \int_{z_1}^{z_2} \kapo \rho(z) \phi[x-u_z(z)] \, dz
\, ,
\label{eq:tzintdef}
\eeq
where 
$\kapo$ is a spatially constant line opacity normalization\footnote{
In the notation of Gayley 1995, the line normalization here is given by 
$ \kappa_{\rm o} v_{\rm th} /\kappa_{\rm e} c =  \left [ {\bar Q} Q_{\rm max}^{- \alpha}/\Gamma(\alpha) \right ]^{1/(1-\alpha)}$,
where $\Gamma(\alpha)$ is the complete Gamma function, and numerical values used here are given in Table 1.
},
and
$u_{\rm z} \equiv \bf{\hat{z}} \cdot \bf{u}$ is the local $z$-projection of the vector flow velocity normalized to the ion thermal speed, ${\bf u} \equiv {\bf \varv}/\vth$.
The line-profile function is taken here to have a normalized gaussian form, $\phi(x) = e^{-x^2}/\sqrt{\pi}$,
with $x = (\nu - \nu_{\rm o})/\Delta \nu_{\rm D}$ the observer-frame frequency displacement from line-center in thermal doppler units $\Delta \nu_{\rm D} = \nu_{\rm o} \vth/c$.

In 1D spherically symmetric models in which variables only depend on the radius $r$,  the ray direction is defined in terms of the local $r$ and a stellar impact parameter $p$, with $|z|=\sqrt{r^2-p^2}$, and the sign taken to be positive (negative) in the forward (backward) hemisphere.
Moreover, since the velocity is purely radial ${\bf u}=u_{\rm r} \bf{\hat{r}}$,  we have simply $u_{\rm z} = \mu_{\rm z} u_{\rm r}$, with radial projection cosine $\mu_{\rm z} = z/r$.

\begin{figure}
    	\begin{minipage}{8.0cm}
      \resizebox{\hsize}{!}  {\includegraphics[]{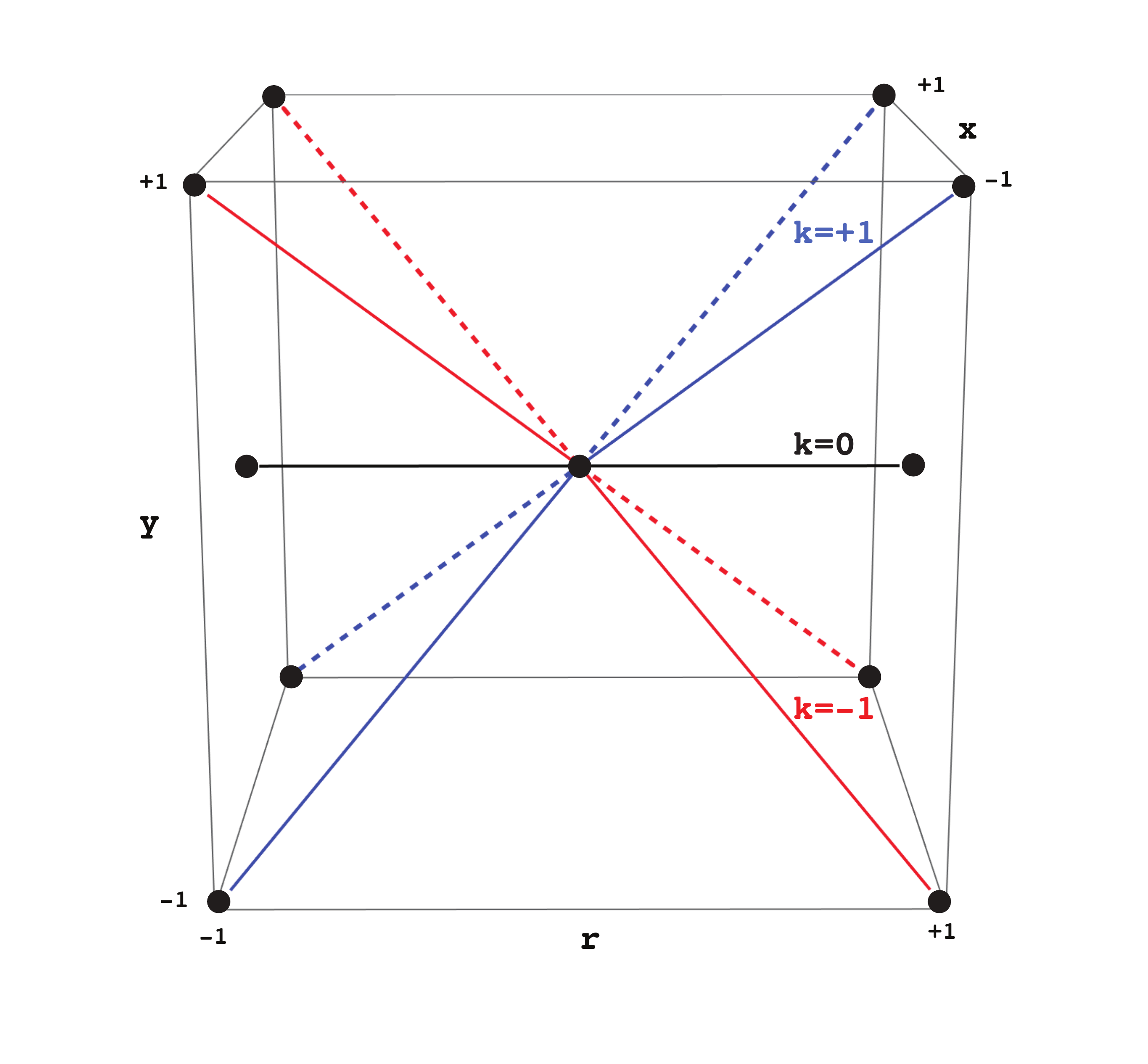}}
    \centering
    \end{minipage}
    	\begin{minipage}{8.0cm}
      \resizebox{\hsize}{!}  {\includegraphics[]{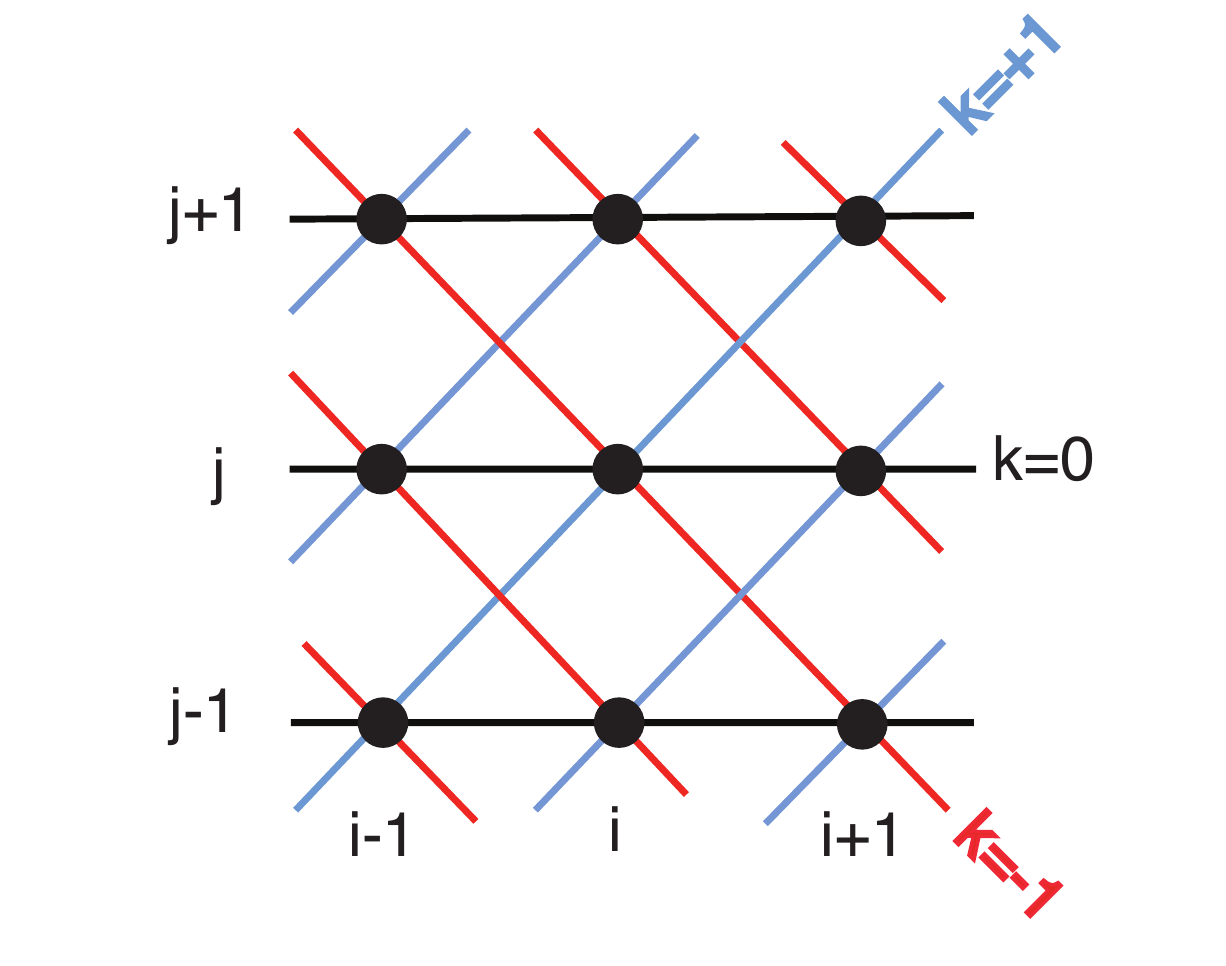}}
    \centering
    \end{minipage}
  \caption{Illustration of ray trajectories in the prograde ($k=+1$; blue), radial ($k=0$; black), and retrograde ($k=-1$; red) directions, crossing grid nodes (black dots) that neighbor a central node with spatial indices $\{i,j\}$. The upper panel shows the full 3D geometry of the radiation rays. But since conditions are assumed constant in x (and thus symmetric about x=0), ray integrations computed along dashed and solid lines of the same color are identical, and so can be accounted for by simply doing one prograde (blue) and one retrograde (red) integration, and then giving these double weight in the angle quadrature. The lower panel shows these final 3 distinct rays projected upon the 2D $r-y$ calculation plane.}  
  \label{Fig:grid}
\end{figure}

In the present 2D pseudo-planar formulation, variations can occur in both radius $r$ and a lateral orthogonal direction $y$, taken to lie in the equatorial plane of symmetry.
The ray directions $z$ now have local projection cosines $\mu_{\rm r}$ and $\mu_{\rm y}$ relative to  the $r$ and $y$ axes, with thus $u_{\rm z} = \mu_{\rm r} u_{\rm r} + \mu_{\rm y} u_{\rm y} = \mu u_{\rm r} + \sin \phi \sqrt{1-\mu^2} u_{\rm y}$, with $\theta \equiv \arccos{\mu}$ and $\phi$ 
the customary radiation angles in \S 2.
Our computations include one purely radial ray, with $\mu_{\rm r}=1$ and $\mu_{\rm y}=0$, so that $u_{\rm z}=u_{\rm r} (r,y)$;
as noted in \S 2, we also formally account for four additional rays that all have $\mu_{\rm r}=1/\sqrt{3}$, 
with two pairs of rays with $\mu_{\rm y}=\pm 1/\sqrt{3}$, but each pair forming mirror projections above/below the $r-y$ plane.
In practice, because of the mirror symmetry about this plane, explicit computation is only needed for one pair, with the other pair simply accounted for by doubling the quadrature weights (see Fig. A1).
For notational convenience, let us denote this triad with an index $k=-1,0,1$, such that $\mu_{{\rm r},0}=1$ and $\mu_{{\rm r},\pm 1} = 1/\sqrt{3}$, while $\mu_{{\rm y},0}=0$ and $\mu_{{\rm y},\pm 1}= \pm 1/\sqrt{3}$ (see Fig. \ref{Fig:grid}).

For our uniform spatial grid with fixed spacings $\Delta r = \Delta y = \Delta = 0.001 \Rstar$, we have coordinates $r_{\rm i}=\Rstar + i \Delta$ and $y_{\rm j} = j \Delta$, for grid indices $i=$1 to $n_{\rm r}=1000$ and $j=$1 to $n_{\rm y}=100$.
At each grid node $\{i,j\}$, the outward (+) increment in optical depth $\Delta t_{\rm +,ijk} (x)$ along each of the directional triad $k$ is computed from (\ref{eq:tzintdef}), assuming a piecewise linear variation of density $\rho$ and velocities $u_{\rm r}$ and $u_{\rm y}$ to the next outer grid node, with indices $\{i+1,j+k\}$.
Summation from the lower boundary at the stellar surface then gives the associated outwardly {\em integrated} optical depths 
along each direction $k$ to some node with coordinates $\{r,y\}$,
\beq
t_{\rm +,k} (x,r,y) = \sum_{\rm i, j}  \Delta t_{\rm +,ijk} (x) + t_{\rm +,k} (x,\Rstar,y_\ast)
\, ,
\label{eq:tpkxry}
\eeq
where the summation is understood to be over all $i$ below the index for $r$, and over the associated $j$ variation for each particular ray $k$;
the assumed periodic variation in $y$ means that  $j$ indices are simply mapped onto the range $0 < j < 100$ by taking ${\rm mod}(j,100)$.
This means that all rays considered here in the pseudo-planar model hit the stellar surface at the lower boundary of the grid. 
The surface boundary value allows one to account for a photospheric line-profile and the effect of a cutoff at a maximum opacity $\kapmax$ in the line distribution, as given by
 equation (OP96-66),
 \beq
 t_{\rm +,k} (x,\Rstar,y_\ast) = \frac{\kapo}{\kapmax} + \frac{\kapo \phi (x)}{\kape}
 \, .
 \label{eq:tpklbc}
 \eeq

With the outward optical depths $t_{\rm +,k} (x,r,y)$ in hand, the computation of the resulting radial component of the line-acceleration follows much the same approach as for the 1D SSF formalism given in section 5.3 of OP96, as further elaborated in section 2.1 of SO15. The {\em direct} absorption component of gravitationally scaled line-acceleration thus takes the form (cf.\ (SO15-3))
\beq
\Gamma_{\rm dir,r} (r,y) =  \Gamma_{\rm thin}  \sum_{x,k} w_{\rm x} w_{\rm k}  \, 
  \phi \left( x- u_{\rm z,k} \right ) 
 t_{\rm +,k}^{-\alpha}  (x,r,y)
\, ,
\label{eq:gamdir}
\eeq
where $u_{\rm z,k} \equiv \mu_{\rm r,k} u_{\rm r} (r,y) + \mu_{\rm y,k} u_{\rm y} (r,y)$,
and the optically thin normalization $\Gamma_{\rm thin}$ is given by (SO15-4).
As in 1D models, the quadrature in frequency $x$ is uniform with equal weights $w_{\rm x} = \Gamma(\alpha)/n_{\rm x}$, and a resolution of three points per thermal doppler width, $\Delta x = 1/3$;
but the 1D angle-quadrature weight $w_{\rm y}$ is now replaced with the triad $w_{\rm k}$, with normalized values $w_0=0.211325$ and and $w_{\pm 1} = 0.394338$. 
This triad weights the oblique rays such that they cover the full $\mu$-space from 0 to $1/\sqrt{3}$, plus half the space from $1/\sqrt{3}$ to a radial ray at $1$;
the radial ray thus get the weight $w_0=(1-1/\sqrt{3})/2 = 0.211235$.

Within this SSF formalism, the {\em diffuse} (scattering) component of the line-force is again (see OP96) formed by using $t_+$ to build an associated {\em inward} optical depth $t_-$, and using this to form a difference between the outward vs. inward escape probability, as given in equation (SO15-6). 
To avoid the variability of a nodal topology at the wind base (see \S 3 of SO15), we assume a simple optically thin source function computed from a uniformly bright surface without limb darkening, as given in equation (SO15-8).

Our computation of the {\em lateral} ($y$) component of the line-force warrants some further elaboration.
While the overall formulation is similar, this now depends
the {\em difference} between the escape probabilities in prograde ($k=+1$) and retrograde ($k=-1$) directions, applied to both the direct and diffuse components. (The radial ray $k=0$ plays no role.)
Defining the profile-averaged, outward  (+) escape probabilities in the prograde/retrograde ($\pm$) directions as
\beq
b_{+,\pm} (r,y) = \sum_{x}   w_x \, \phi \left( x- u_{z,\pm1}) \right )  t_{+,\pm 1}^{-\alpha}  (x,r,y)
\, ,
\label{eq:bppmdef}
\eeq
we can write the direct component of the lateral line-acceleration (still scaled by the radial gravity) as
\beq
\Gamma_{\rm dir,y} (r,y) = \Gamma_{\rm thin} f(r) \left [ b_{+,+} (r,y) -  b_{+,-} (r,y) \right ] 
\, .
\label{eq:gamdiry}
\eeq
To account for the additional radial drop-off associated with angular shrinking of the stellar core \citep[e.g.,][]{Gayley00}, 
we include here a correction factor 
\beq
f(r) = w_1 \mu_{\rm y,+1} \,  \frac{\Rstar^2}{r^2} 
\, .
\label{eq:fdef}
\eeq

Defining {\em inward} (--) escape probabilities in a way analogous to (\ref{eq:bppmdef}), we can write the associated {\em diffuse} component of the lateral line-acceleration as
\beq
\Gamma_{\rm diff,y} (r,y) = \Gamma_{\rm thin} \, s(r) \left [ b_{-,+} +  b_{+,-}  - b_{+,+}  - b_{-,-}  \right ]
\, ,
\label{eq:gamdiffy}
\eeq
where the optically thin source function factor $s(r)$ is given by equation (SO15-8).
For both the radial ($r$) and lateral ($y$) components, the associated total acceleration is given by the sum of the direct and diffuse contributions, $\Gamma_{\rm tot} = \Gamma_{\rm dir} + \Gamma_{\rm diff}$.

In our numerical radiation-hydrodynamics simulations, we apply these total radial and lateral line-accelerations  in the associated radial and lateral momentum equations,
\beqa
\frac{\partial \varv_r}{\partial t} + \varv_r \frac{\partial \varv_r}{\partial r} = -\frac{1}{\rho} \frac{dP}{dr}  &+& (\Gamma_{\rm tot,r}-1) \frac{GM_{\rm eff}}{r^2} + \frac{\varv_{\rm y}^2}{r}
\,
\label{eq:momr}
\\
\frac{\partial \varv_y}{\partial t} + \varv_y \frac{\partial \varv_y}{\partial r} =  -\frac{1}{\rho} \frac{dP}{dy}  &+& \Gamma_{\rm tot,y} \frac{GM_{\rm eff}}{r^2}  - \frac{\varv_{\rm r} \varv_{\rm y}}{r}
\, ,
\label{eq:momt}
\eeqa
where $P$ is the gas pressure, $M_{\rm eff} \equiv M(1-\Game)$, and the last term in each equation corrects our pseudo-planar
treatment for curvilinear coordinate effects (`centrifugal' 
and `coriolis' forces) in a spherical outflow.
The density $\rho$ is evolved according to the mass continuity equation,
\beq
\frac{\partial \rho}{\partial t} + \frac{\partial (\rho \varv_r)}{\partial r} + \frac{\partial (\rho \varv_y)}{\partial y} = - \frac{2 \rho \varv_r}{r}
\, ,
\label{eq:masscon}
\eeq
where the source term on the right-hand-side corrects for the neglect of the spherical divergence within our pseudo-planar treatment of the flow divergence $\nabla \cdot (\rho {\bf \varv})$.
The $1/r^2$ decline of the radiative flux, which sets the scale of the line-accelerations, is accounted for by scaling these accelerations with the inverse-square decline of the stellar gravity.

\end{document}